\documentstyle[preprint,aps,version2]{revtex}
\tightenlines
\begin{document}
\draft

\pagestyle{empty}

\newcommand{\rb}{$\bf {R_{\mbox{$\bf b$}}}$}
%\newcommand{\postscript}[2]
%{\setlength{\epsfxsize}{#2\hsize}
%\centerline{\epsfbox{#1}}}

\preprint{
\noindent
\begin{minipage}[t]{3in}
\begin{flushleft}
UCB--PTH--95/40 \\
LBL--37932 \\
LMU--TPW--95--18
\end{flushleft}
\end{minipage}
\hfill
\begin{minipage}[t]{3in}
\begin{flushright}
SLAC--PUB--95--7050 \\
NSP--ITP--95--150 \\
hep-ph/9511324 \\
November 1995
\end{flushright}
\end{minipage}
}

\vspace*{-0.3in}

\begin{title}
The Light Higgsino-Gaugino Window
\end{title}

\author{
\vspace*{-0.15in}
Jonathan L. Feng
\thanks{Research Fellow, Miller Institute for Basic Research in
Science.}
\thanks{Work supported in part by the Director, Office of Energy
Research, Office of High Energy and Nuclear Physics, Division of High
Energy Physics of the U.S. Department of Energy under Contract
DE--AC03--76SF00098 and in part by the National Science Foundation under
grant PHY--90--21139.}
}

\begin{instit}
Theoretical Physics Group, Lawrence Berkeley National Laboratory\\
University of California, Berkeley, California 94720\\
and\\
Department of Physics\\
University of California, Berkeley, California 94720
\end{instit}

\author{
\vspace*{-0.15in}
Nir Polonsky
\thanks{Supported by a grant of the DFG.}
}

\begin{instit}
Sektion Physik (Lehrstuhl Prof. Wess), Universit\"at M\"unchen\\
37 Theresienstrasse, D--80333 M\"unchen, Germany
\end{instit}

\author{
\vspace*{-0.15in}
Scott Thomas
\thanks{Work supported by the Department of Energy under Contract
DE--AC03--76SF00515 and by the National Science Foundation under grant
PHY--94--07194.}
}

\begin{instit}
Stanford Linear Accelerator Center\\
Stanford University, Stanford, California 94309\\
and\\
Institute for Theoretical Physics\\
University of California, Santa Barbara, CA 93106
\end{instit}

%\receipt{}

\vspace*{-0.43in}
\begin{abstract}
Supersymmetric models are typically taken to have $\mu$ parameter and
all soft supersymmetry breaking parameters at or near the weak scale.
We point out that a small window of allowed values exists in which $\mu$
and the electroweak gaugino masses are in the few GeV range.  Such
models naturally solve the supersymmetry $CP$ problem, can reduce the
discrepancy in $R_b$, and suppress proton decay.  In this window two
neutralinos are in the few GeV range, two are roughly degenerate with
the $Z^0$, and both charginos are roughly degenerate with the $W^{\pm}$
bosons.  Such a signature cannot escape detection at LEP II.  Models
that fall in this window automatically arise from renormalizable hidden
sectors in which hidden sector singlets participate only radiatively in
supersymmetry breaking.
%\centerline{(Submitted to . . .)}
\end{abstract}

%\pacs{}

\newpage
\pagestyle{plain}
\narrowtext

\setcounter{footnote}{0}

\section{Introduction}

It is usually assumed that supersymmetric model are required to have
dimensionful parameters $\mu, M_{\mbox{\tiny SUSY}} \gtrsim {\cal
O}(M_W)$, where $\mu$ is the supersymmetric Higgs mass, and
$M_{\mbox{\tiny SUSY}}$ is the scale of the visible sector supersymmetry
(SUSY) breaking parameters. We note here that viable models exist in
which $\mu$ and the electroweak gaugino masses are ${\cal
O}(\mbox{GeV})$.  Contrary to common lore, such parameters are allowed
despite many stringent constraints, including those arising from the LEP
$Z^0$ width measurements.

Although the allowed parameter space is not large, such models have a
number of interesting features: the SUSY $CP$ problem is solved, the
current discrepancy between theoretical and experimental values of $R_b$
can be reduced, and proton decay is suppressed.  In addition, we will
demonstrate that satisfactory electroweak symmetry breaking may be
achieved and discuss how such models might arise in supergravity
theories from a renormalizable hidden sector.  An important and
unambiguous prediction of such models is the observation of neutralinos
and charginos at LEP II.

\section{$Z^0$ Width Constraints}

We first discuss the bounds from $Z^0$ decays.  Because the $\mu$
parameter and electroweak gaugino masses enter chargino and neutralino
mass matrices, one might expect that when these parameters are in the
GeV range, charginos and neutralinos are light and in conflict with the
bounds on $Z^0$ decay widths.  We will see, however, that in some of
this region of parameter space, charginos are sufficiently massive and
neutralinos are sufficiently decoupled from the $Z^0$ that these bounds
may be satisfied.

First consider the charginos.  We assume that the charginos and
neutralinos are the standard mixtures of electroweak gauginos and the
Higgsinos of the two Higgs doublets.  We also denote the bino, wino, and
gluino masses by $M_1$, $M_2$, and $M_3$, respectively, and the ratio of
Higgs expectation values by $\tan\beta \equiv \langle H_2 \rangle /
\langle H_1 \rangle $.  The chargino mass terms are then $(\psi^-)^T
{\bf M}_{\tilde{\chi}^{\pm}} \psi^+ + \mbox{ h.c.}$, where the mass
matrix is

\begin{equation}\label{chamass}
{\bf M}_{\tilde{\chi}^{\pm}} = \left( \begin{array}{cc}
M_2                   & \sqrt{2} \, M_W\sin\beta \\
\sqrt{2} \, M_W\cos\beta   & \mu                  \end{array}
\right)
\end{equation}
in the basis $\psi^{\pm} = (-i\tilde{W}^{\pm}, \tilde{H}^{\pm})$.  The
current bound on chargino masses from LEP measurements is 47 GeV
\cite{PDG}.  This bound requires no additional assumptions,
as charginos remain coupled to the $Z^0$ for all values of the
parameters.  We see, however, that for $\mu, M_2 \approx 0$ and
$\tan\beta \approx 1$, both charginos have mass $M_W$ and avoid the
bound.  With $\mu \approx 0$ ($M_2 \approx 0$), as $M_2$ ($\mu$)
increases, one chargino mass eigenvalue drops by the see-saw mechanism,
and when $M_2$ ($\mu$) $> 90 \mbox{ GeV}$, the chargino mass limit is
violated for all $\tan\beta$.  However, for $1 < \tan\beta \lesssim
2.1$, the parameters $\mu, M_2 \approx 0$ satisfy the chargino mass
bound.

Next we examine the neutralino sector.  Unlike charginos, neutralinos
may completely decouple from the $Z^0$, and for this reason, there are
no strict lower bounds on neutralino masses.  If one assumes gaugino
mass unification and $\tan\beta > 2$, the lower bound on the lightest
neutralino's mass is 20 GeV \cite{PDG,L3,otherLEP}.  For $\tan\beta
\lesssim 1.6$, however, this mass bound disappears altogether
\cite{L3,otherLEP}.  It is clear, then, that a discussion of light
neutralinos requires a detailed analysis of their couplings to the $Z^0$
boson.  The $Z^0$ width constraints are therefore considerably more
complicated for neutralinos than for charginos, and we will discuss them
in two stages.  First, we present a simple discussion that makes clear
the qualitative features of the allowed region.  These features are
illustrated in Fig.~\ref{fig:1}.  We then add a number of refinements to
the analysis and present the resulting allowed region in
Fig.~\ref{fig:2}.

It is convenient to write the neutralino mass terms
$\frac{1}{2} (\psi ^0)^T {\bf M}_{\tilde{\chi}^0} \psi^0 + \mbox{ h.c.}$
in the basis $(\psi^0)^T = \left(
\frac{1}{\sqrt{2}} ( -i \tilde{Z^{0}} + \tilde{H}_A ),
\frac{1}{\sqrt{2}} ( -i \tilde{Z^{0}} - \tilde{H}_A ),
-i\tilde{\gamma}, \tilde{H}_S \right)$,
where $\tilde{H}_A \equiv \tilde{H}_1 \cos\beta - \tilde{H}_2
\sin\beta$, and $\tilde{H}_S \equiv \tilde{H}_1 \sin\beta + \tilde{H}_2
\cos\beta$.  The tree level mass matrix is then

\begin{equation}\label{neutmass}
{\bf M}_{\tilde{\chi}^0} = \left(
\begin{array}{cccc}
M_Z+\frac{1}{2}M+\frac{1}{2}\mu\sin 2\beta &
\frac{1}{2}M-\frac{1}{2}\mu\sin 2\beta &
\Delta M & -\frac{1}{\sqrt{2}}\mu \cos 2\beta \\
\frac{1}{2}M-\frac{1}{2}\mu \sin 2\beta &
-M_Z+\frac{1}{2}M+\frac{1}{2}\mu\sin 2\beta &
\Delta M & \frac{1}{\sqrt{2}}\mu \cos 2\beta \\
\Delta M & \Delta M & M_{\tilde{\gamma}} & 0 \\
-\frac{1}{\sqrt{2}}\mu\cos 2\beta & \frac{1}{\sqrt{2}}\mu\cos 2\beta
& 0 & -\mu \sin 2\beta        \end{array}
\right) \ ,
\end{equation}
where $M \equiv M_1 \sin^2 \theta_W + M_2 \cos^2 \theta_W$,
$M_{\tilde{\gamma}} \equiv M_1 \cos^2 \theta_W + M_2 \sin^2 \theta_W$,
$\Delta M \equiv \frac{1}{\sqrt{2}} (M_2-M_1)
\cos\theta_W\sin\theta_W$, and $\theta_W$ is the weak mixing angle.
As discussed above, the chargino mass bound is satisfied with $\mu , M_2
\approx 0$. In order to avoid a light neutralino with unsuppressed
coupling to the $Z^0$, it is also necessary that $M_1$ be small.  In the
limit $\mu, M_2, M_1 \to 0$, the basis states given above are mass
eigenstates, with masses $M_Z$, $M_Z$, 0, and 0.  The light photino,
$\tilde{\gamma}$, does not couple to the $Z^0$, and the light Higgsino,
$\tilde {H}_S$, decouples for $\tan \beta \to 1$.  In this case, the
only nonzero coupling of the neutralinos to the $Z^0$ is through
$Z^{0}\tilde{H}_A\tilde{H}_S$, which is suppressed by phase space.

To understand how far one can vary from the limit $\mu, M_2, M_1, \tan
\beta -1 \to 0$ and still satisfy all the constraints, we must discuss
the bounds in greater detail.  Let us denote the lightest neutralino,
$\tilde{\chi}^0_1$, by $\chi$, and the heavier neutralinos,
$\tilde{\chi}^0_2$, $\tilde{\chi}^0_3$, and $\tilde{\chi}^0_4$, by
$\chi'$.  We will assume that the lightest neutralino $\tilde{\chi}^0_1$
is the lightest supersymmetric particle (LSP) and escapes the detector.
There are then bounds on $\Gamma(Z^0\to \chi\chi)$ from the invisible
$Z^0$ width, and bounds on $\Gamma(Z^0\to \chi\chi')$ and $\Gamma(Z^0\to
\chi'\chi')$ from direct searches for neutralinos.

The current bound on the invisible width of the $Z^0$, in units of the
neutrino width, is $N_{\nu} = 2.988 \pm 0.023$ \cite{PDG,EWG}. The
2$\sigma$ upper bound on non-Standard Model invisible decays is then
$\delta N_{\nu} = 0.034$, or a $Z^0$ branching ratio of $B_{\mbox{\tiny
inv}} = 2.3 \times 10^{-3}$.  We will take this as the bound on the
$\chi\chi$ width.\footnote{Formally, production of $\chi\chi'$ and
$\chi'\chi'$, if followed by $\chi'\to \chi\nu\bar{\nu}$, will also
contribute to the invisible width.  However, as we will see, such
processes violate the visible width bounds long before their effect on
the invisible width becomes important, and so may be safely ignored
here.}

The visible width bounds are determined from direct searches for
neutralinos.  In Ref.~\cite{L3} the L3 Collaboration placed bounds on
neutralinos based on an event sample including 1.8 million hadronic
$Z^0$ events.  The decays $\chi'\to \chi {Z^0}^* \to \chi f \bar{f}$,
with $f = q, e, \mu$, and also the radiative decay $\chi' \to \chi
\gamma$ were considered.  For given masses $m_{\chi}$ and $m_{\chi'}$,
neutralino events were simulated, and the photonic branching ratio was
chosen to give the weakest bounds.  In the regions of most interest to
us, the neutralino masses are $m_{\chi} \approx 0$ and $m_{\chi'} \approx 0,
M_Z$.  For these masses, the upper bound on the branching ratio
$B(Z^{0}\to\chi\chi')$ ($B(Z^{0}\to\chi'\chi')$) was found to be at
least $1.2\times 10^{-5}$ ($3.5\times 10^{-5}$).  These bounds
deteriorate rapidly as $m_{\chi} \to m_{\chi'}$, but we will
conservatively assume that they apply for all masses.

Given these bounds, we may now determine the allowed region of parameter
space. As one varies from the point $\mu, M_2, M_1, \tan\beta-1 = 0$,
the basis states begin to mix, and have masses given by the diagonal
elements up to corrections of ${\cal O}((M_1, M_2, \mu)^2/M_Z)$.  The
mixing angles between the heavy states and between the heavy and light
states are ${\cal O}((M_1, M_2, \mu)/M_Z)$.  However, (at tree level)
the mixing between the light states occurs only through the intermediate
heavy states and so is ${\cal O}((M_1, M_2, \mu)/M_Z)^2$. These mass
shifts and mixings may then weaken the various coupling constant and
phase space suppressions.  Decays to the following three states
determine the allowed region:

\noindent (a) $\tilde{H}_S \tilde{H}_S$. The ratio $\Gamma(Z^0 \to
\tilde{H}_S \tilde{H}_S)/\Gamma(Z^0 \to \nu\bar{\nu})$ is $\cos^2 2\beta$.
If $\tilde{\chi}^0_2$ has a significant $\tilde{H}_S$ component, the
stringent limits on the visible $Z^0$ width require $\tan \beta < 1.02$,
a range that is in conflict with the perturbativity of the top Yukawa
coupling (see below).  However, when $\tilde{\chi}^0_1 \approx
\tilde{H}_S$, the constraint on $\tan\beta$ comes only from the
invisible width bound, which is two orders of magnitude weaker.  The LSP
(at tree level) is very nearly pure $\tilde{H}_S$ for
$|M_{\tilde{\gamma}}|>|\mu|\sin 2\beta$ and satisfies the invisible
width bound for $\tan \beta < 1.20$.  Below, we assume that this
constraint is satisfied and $\tilde{\chi}^0_1 \approx \tilde{H}_S$.
This scenario was previously considered in Ref.~\cite{photinodecay}.

\noindent (b) $\tilde{H}_S \tilde{\chi}^0_3$.  The decay to $\tilde{H}_S
\tilde{\chi}^0_3$ is suppressed only by phase space.  This suppression
is adequate when both $M_Z+\frac{1}{2}M+\frac{1}{2}\mu\sin 2\beta +
|\mu| \sin 2\beta \gtrsim M_Z$ and $M_Z-\frac{1}{2}M -
\frac{1}{2}\mu\sin 2\beta + |\mu| \sin 2\beta \gtrsim M_Z$.  Assuming
$M>0$, this constraint is then $M \lesssim 3|\mu|\sin 2\beta$ for
$\mu<0$ and $M\lesssim |\mu|\sin 2\beta$ for $\mu>0$.\footnote{For
$M<0$, the requirements are $|M|<|\mu|\sin 2\beta$ for $\mu<0$ and
$|M|<3|\mu|\sin 2\beta$ for $\mu>0$.  However, we will concentrate on
the case $M>0$, as this holds in most of the allowed region.}

\noindent (c) $\tilde{H}_S \tilde{\chi}^0_2$.  For this decay to be
suppressed, the neutralino $\tilde{\chi}^0_2$ must be nearly a pure
photino.  The mixing of this eigenstate is controlled by $\Delta M$ and
vanishes when $\Delta M=0$, that is, when $M_1=M_2$.

The allowed regions for $\tan\beta = 1.15$ are presented in
Fig.~\ref{fig:1} for three values of the ratio $M_1/M_2$.  The allowed
regions are very similar for all $1.02 < \tan\beta < 1.20$.  Constraints
(a) and (b) limit the allowed parameter space to a region with
boundaries of definite slope as given above.  Constraint (c) provides a
maximum allowed $M_2$, and, as expected, disappears in the limit
$M_1=M_2$.

The analysis above gives a rough picture of what parameter regions may
survive the various constraints.  However, several refinements are
necessary.  First, as noted above, the photonic branching ratio was
assumed to be unknown in the analysis of Ref.~\cite{L3} and was chosen
to give the weakest bounds.  However, for specific branching ratios,
additional regions might be excluded.  In particular, the radiative
photon decay $\tilde{\chi}^0_2 \to \tilde{\chi}^0_1 \gamma$ has been
studied previously \cite{photinodecay} and is expected to be dominant in
our case, where $\tilde{\chi}^0_2 \approx \tilde{\gamma}$ and
$\tilde{\chi}^0_1 \approx \tilde{H}_S$.  The production of
$\tilde{\chi}^0_1 \tilde{\chi}^0_2$ then gives a spectacular single
photon signal, and the bound on its rate can be significantly improved.
To estimate this new bound, we reexamine the data of Ref.~\cite{L3}.  In
that event sample, the dominant Standard Model background, $Z^0 \to
\gamma_{\mbox{\tiny ISR}}\nu\bar{\nu}$, is expected to produce only
$15.7 \pm 1.5$ events with photons passing the cut $p_T > 10 \mbox{
GeV}$.  Assuming that $\sqrt{15.7} \simeq 4$ neutralino events could be
hidden in this background, that the efficiency of neutralino detection
in this mode is 50\%, as given in Ref.~\cite{L3}, and, for simplicity,
that the neutralino events are uniformly distributed in the range
$0<p_T<45 \mbox{ GeV}$, we find an upper bound of 10 signal events.  We
therefore consider the effect of strengthening the bound to
$B(Z^0\to\chi\chi') < 3.9\times 10^{-6}$ (we also take
$B(Z^0\to\chi'\chi') < 3.9\times 10^{-6}$).\footnote{This estimate is in
excellent agreement with the bound of $4.3\times 10^{-6}$ set by the
OPAL Collaboration\cite{singlephoton} on exotic decays $Z^0\to X\gamma$,
where $X$ decays invisibly.}

Another important refinement is to include the data taken above $M_Z$.
As the process $Z^0 \to \tilde{H}_S \tilde{\chi}_3^0$ is suppressed only
by phase space, the boundary of the allowed region defined by this
constraint can be expected to be very sensitive to deviations in
$\sqrt{s}$ from $M_Z$.  The analysis of Ref.~\cite{L3} used 1993 data,
which included an integrated luminosity of 18 $\mbox{pb}^{-1}$ at
$\sqrt{s}= M_Z+1.8 \mbox{ GeV}$ \cite{EWG}.  This data sample then
includes approximately 240,000 hadronic $Z^0$ decays, and we estimate
that with this much data the branching ratio bounds are degraded by a
statistical factor of 2.7 at the higher energy.

In Fig.~\ref{fig:2}, we plot the new allowed region, including the
tighter branching ratio bound from radiative neutralino decay and the
effects of data taken above $M_Z$.  Points in the allowed region are
values of $\mu$ and $M_2$ that are allowed for some $M_1$ in the range
$\frac{1}{2} M_2 \le M_1 \le 2 M_2$.  (This range has been chosen rather
arbitrarily.  By considering $M_1> 2 M_2$, the allowed region can be
extended to lower values of $M_2$ in the negative $\mu$ region.)
Qualitatively, the allowed region is very similar to what would be
expected from Fig.~\ref{fig:1}, with the exception that points with $\mu
\gtrsim 0$ have been eliminated.  These points required the phase space
suppression of $\tilde{H}_S \tilde{\chi}^0_3$ production, and are
eliminated by the $M_Z + 1.8$ GeV data.  We see, however, that much of
the $\mu<0$ region still remains.

In the previous figures, radiative corrections have not been included.
Radiative corrections to the diagonal entries of Eq.~(\ref{neutmass})
shift the neutralino masses, and thus shift the boundaries slightly.
Corrections to the off-diagonal entries introduce mixings between
states.  Mixings between the heavy states and between heavy and light
states are unimportant as similar mixings are already present at tree
level, and all such mixings are highly suppressed because they mix
states whose eigenvalues are split by ${\cal O}(M_Z)$.  Off-diagonal
radiative corrections that mix the light states can be important,
however, as they give a Dirac mass that lifts the tree level zero in
Eq.~(\ref{neutmass}) \cite{HallRandall}.  The largest such correction
comes from top-stop loops (similar to the ones that induce the radiative
photon decay), is of order 1 GeV, and vanishes when the left- and
right-handed stops, $\tilde{t}_L$ and $\tilde{t}_R$, are degenerate
\cite{HallRandall}.  When this radiative mixing of the light states is
significant with respect to the tree level masses, the LSP is a mixture
of both $\tilde{H}_S$ and $\tilde{\gamma}$.  Some regions of the window
that were allowed at tree level are then excluded by the stringent $Z^0$
visible width bound.  However, for $\mu$ and $M_2$ sufficiently large,
the radiative mixing becomes negligible, and the LSP can be mostly
$\tilde{H}_S$.  For stop masses in the range $100 \mbox{ GeV} <
m_{\tilde{t}_R}, m_{\tilde{t}_L} < 300 \mbox{ GeV}$, we find that the
effect of the radiative Dirac mass is to remove points with $M_2, \mu
\lesssim 2-4 \mbox{ GeV}$, leaving most of the allowed region displayed
in Fig.~\ref{fig:2} intact.

Returning to the chargino mass matrix of Eq.~(\ref{chamass}), we see
that, in the allowed Higgsino-gaugino window, both charginos are roughly
degenerate with the $W^{\pm}$.  Charginos and neutralinos are thus all
within reach of LEP II; if observed there, precision studies may be able
to determine if the SUSY parameters lie in this allowed window
\cite{FS}. The chargino mass is lowered by the deviation from $\mu, M_2,
\tan\beta-1 = 0$, but remains above 70 GeV.  It is important to note
that $m_{\tilde{\chi}^{\pm}_1} + m_{\tilde{\chi}^0_1} > 77 \mbox{ GeV}$
throughout almost all of the allowed region and grows beyond $M_W$ as
$\mu$ and $M_2$ increase in the allowed region, so the branching ratio
for the decay $W \to \tilde{\chi}^{\pm}_1 \tilde{\chi}^0_1$ is highly
suppressed \cite{Tata}.

Another possible constraint on the light Higgsino-gaugino window is from
the relic dark matter density.  As the LSP is mostly Higgsino, the
dominant annihilation channel is through $s$-channel $Z^0$ to light
fermion pairs.  The Higgsino-$Z^0$ coupling is necessarily suppressed in
order to avoid the invisible width bound from $Z^0$ decay.  This
generally leads to an overproduction of primordial Higgsinos.  Using the
non-relativistic approximation for the freeze out density and the
annihilation cross section for a pure Higgsino state given in
Ref.~\cite{dmref}, we find that for $\tan \beta \simeq 1.2$, $\Omega h^2
\lesssim 1$ only for $m_{LSP} \gtrsim 20$ GeV.  We therefore conclude
that either there is an additional entropy release below the LSP freeze
out temperature to dilute the relic Higgsinos, or that $R$ parity is
broken so that the LSP is not stable, and therefore does not contribute
to the dark matter.

\section{Radiative Symmetry Breaking}

Before discussing the scalar sector and radiative symmetry breaking, we
note that, in the allowed window where $\mu, M_1, M_2 \ll M_W$, an
approximate $U(1)$ $R$-symmetry exists in the weak gaugino and Higgs
sector, under which $R(H_1)=R(H_2)=0$, and all other chiral matter
fields have $R=1$.\footnote{Under an $R$ transformation the scalar,
fermionic, and auxilliary components of a chiral superfield transform as
$\phi \rightarrow e^{i \alpha R} \phi$, $\psi \rightarrow e^{i \alpha
(R-1)} \psi$, and $F \rightarrow e^{i \alpha (R-2)} F$, respectively,
where $R$ is the superfield's $R$ charge.  The superpotential has $R$
charge $R(W)=2$, and a gauge superfield has $R$ charge
$R(W^{\alpha})=1$.}  It is possible to promote this approximate symmetry
to an exact symmetry of the entire MSSM Lagrangian, in which case all
gaugino masses, $A$ terms, and $\mu$ would vanish
\cite{HallRandall,RR,RS,bw}. We do not impose such a symmetry by hand,
but simply note that an approximate symmetry exists in the allowed
window.  Below we discuss some consequences of extending the approximate
$U(1)_R$ symmetry to other sectors of the MSSM.  Such an approximate
symmetry in fact arises accidentally in certain types of hidden sector
SUSY breaking scenarios as discussed below.

The allowed window requires $\tan \beta \approx 1$.  Let us therefore
reexamine the lower bound on $\tan \beta$ from the requirement that the
top Yukawa coupling, $h_t$, remain perturbative to high scales.  This is
related to the top quark pole mass by the one-loop relation
\cite{Damien}

\begin{equation}
h_{t}(m_{t}) \simeq \frac{m_{t}^{\mbox{\tiny pole}}}{174{\mbox{ GeV}}}
\frac{\sqrt{1 + \tan^{2}\beta}}{\tan\beta}
\left[ 1 -\frac{5}{3}\frac{\alpha_{s}}{\pi} -
\Delta_{\mbox{\tiny SUSY QCD}} - \Delta_{\mbox{\tiny electroweak}}
\right] \lesssim 1.15 \ ,
\label{ht}
\end{equation}
where $1.15$ is our estimate of the quasi-fixed point value.  Neglecting
SUSY, electroweak, and higher loop corrections, one has a $\sim 6\%$
correction to the tree level result, and taking $m_{t}^{\mbox{\tiny
pole}} \gtrsim 160$ GeV, we find the constraint $\tan\beta \gtrsim
1.14$.  However, including the one-loop SUSY QCD corrections
\cite{Damien,donini}, we find an additional few percent correction (for a
nonvanishing gluino mass) whose sign depends on the various SUSY
parameters.  A $\sim 10\%$ correction is thus possible, which would
lower the $\tan\beta$ bound to $\tan\beta \gtrsim 1.04$.  Alternatively,
for a fixed $\tan\beta$ the perturbativity upper bound on
$m_{t}^{\mbox{\tiny pole}}$ could increase.  Thus, one can still
consider perturbative values of $h_t$ at Planckian scales for $\tan\beta
\lesssim 1.2$, and we may also consider the possibility of radiative
symmetry breaking (RSB).

Next we consider the scalar Higgs sector.  In the allowed window, the
Higgs potential is $V = m_{1}^{2}H_{1}^{2} + m_{2}^{2}H_{2}^{2} -
m_{12}^{2}(H_{1}H_{2} + \mbox{ h.c.})+ D$--terms + $\Delta
V^{\mbox{\tiny 1-loop}}$, where $m_i$ are in our case simply the soft
SUSY breaking masses, since $\mu$ is generally small.  The condition
$m_{2}^{2} < 0$ triggers electroweak symmetry breaking, and $m_{1}^{2}>
0$ is required for the potential to be bounded.  At tree level, the
pseudoscalar Higgs mass is $m_{A}^2 = m_{12}^2 (\tan\beta + \cot\beta
)$.  Although often assumed, it is not generally true in supergravity
theories that $m_{12}^2$ is proportional to $\mu$.  A small $\mu$
parameter does not, therefore, imply the existence of a light
pseudoscalar.  Note also that a large $m_{12}^2$ does not violate the
approximate $U(1)_R$ symmetry given above.

We have seen above that $\tan\beta \approx 1$ in the allowed window.  At
tree level, the light $CP$ even Higgs mass satisfies the bound
$m_{h^{0}} < M_Z\cos 2\beta$ and so vanishes in the limit $\tan \beta
\to 1$.  However, $m_{h^{0}} \propto h_{t}m_{t}$ is generated by
top-stop contributions to $\Delta V^{\mbox{\tiny 1-loop}}$, and, in
principle, a large Higgs mass can be obtained to satisfy the current
experimental bound of $m_{h^{0}} \gtrsim 60$ GeV
\cite{PDG}.  (The lower bound on the Standard Model Higgs boson
mass is the relevant one in the limit $\tan\beta \rightarrow 1$.)  If
the approximate $U(1)_R$ symmetry is extended to the entire Lagrangian,
though, achieving $m_{h^{0}} \gtrsim 60$ GeV is not trivial.  In this
case the mixing between the stops $\tilde{t}_L$ and $\tilde{t}_R$, which
can significantly enhance the loop contributions to $m_{h^0}$ \cite{LP},
is small since $\mu, A \approx 0$.  In addition, the stop masses
$m_{\tilde{t}_{L,\,R}}$ are constrained from above if RSB with minimal
particle content is required.  This may be seen by recalling the
minimization condition $m_2^2 = \left(m_1^2 + \frac{1}{2}M_Z^2
(1-\tan^2\beta)\right)/\tan^2\beta$.  The constraint $m_1^2>0$ then
implies a lower bound on $m_{2}^{2}$ of $-
\frac{1}{2}M_Z^2 (\tan^2\beta - 1)/\tan^2\beta$.  In the $U(1)_R$
symmetric case, where all gaugino masses are small, the RGE equation for
$m^2_2$ is $\partial{m_{2}^{2}}/\partial{\ln{Q}} \simeq
\frac{3}{8\pi^{2}} h_{t}^{2}[m_{2}^{2} + m^2_{{\tilde{t}}_{L}}+
m^2_{{\tilde{t}}_{R}}]$. The requirement that $m_2^2$ not be driven too
negative then places an upper bound of typically $\lesssim M_Z$ on the
boundary condition for the stop masses at the grand scale.  One then
finds that the stop masses at the weak scale are not large enough to
push $m_{h^0}$ above its lower bound.  Thus, unless the $U(1)_R$
symmetry is explicitly broken by a gluino mass, RSB and $m_{h^{0}}
\gtrsim 60$ GeV cannot be achieved simultaneously with minimal particle
content. (Note that in Ref.~\cite{RR} the authors assume a global
$U(1)_R$ symmetry in the whole Lagrangian, but do not require
satisfactory RSB.)

Let us elaborate on the above observations. We have seen that to have
satisfactory RSB with minimal particle content, the combination
$[m_{2}^{2} + m_{\tilde{t}_{L}}^{2} +m_{\tilde{t}_{R}}^{2}]$ that
controls $m_{2}^{2}$ renormalization is constrained to be approximately
zero at the grand scale.  If we assume also a common scalar mass $m_{0}$
at the grand scale and vanishing gaugino masses and $A$ parameters
($i.e.$, the $U(1)_R$ symmetric limit), one finds $m_{0} \lesssim
\frac{1}{3} M_{Z}$ (for $\tan\beta \lesssim 1.15$ and $h_{t}$ at its
quasi-fixed point) and an unacceptable spectrum.  (Stronger constraints
apply for non-vanishing dimension-three terms, and the symmetry limit is
preferred.)  However, if we relax the universality assumption, we are
led to consider the following soft parameter boundary conditions at the
grand scale: $m_{2}^{2}(0) \approx -[m_{\tilde{t}_{L}}^{2}(0)
+m_{\tilde{t}_{R}}^{2}(0)] > 0$ and $m_{2}^{2}(0) \neq m_{1}^{2}(0)$.
If we add a non-vanishing gluino mass, thereby explicitly breaking the
$U(1)_R$ symmetry in the colored sector, $m^2_{\tilde{t}_{L}}$ and
$m^2_{\tilde{t}_{R}}$ both turn positive and possibly large in the
course of renormalization, and the radiatively induced $m_{h^{0}}$ is
sufficiently large.  (Such boundary conditions can be realized, $e.g.$,
in certain stringy schemes \cite{stringy}.)  As long as the scalar
potential in the full and effective theories is bounded from below at
all scales, these boundary conditions are acceptable. In particular, we
find solutions with right-handed stops ranging in mass from 45 GeV to
many hundreds of GeV, Higgs bosons in the 60--70 GeV range, and the two
charginos between 70--90 GeV (as is favored by $R_b$ (see below)).  We
present typical spectra in Fig.~\ref{fig:3}, assuming the above pattern
for boundary conditions.  Only those masses that are constrained by RSB
and $m_{h^0}$ are presented.  Note that because the trilinear terms in
the scalar potential are small, dangerous color breaking directions of
the potential, which are generic in the limit $\tan\beta \to 1$
\cite{LP}, are eliminated.  (However, if $m_{\tilde{t}}\lesssim m_t$,
dangerous directions may persist.)

The above scheme is an example of boundary conditions that can
successfully generate RSB.  (Note that all other boundary conditions are
only negligibly constrained by RSB and $m_{h^{0}}$.)  The tuning
required in order to achieve RSB is a reflection of the fact that we did
not have at our disposal an arbitrary $\mu$, which typically absorbs the
tuning. (For example, generically one expects $\mu \gtrsim 1$ TeV for
$\tan\beta \to 1$ \cite{LP}.)  Instead, the tuning is now in the soft
parameters.  Alternatively, $\partial m_{2}^{2}/\partial
\ln{Q}$ can be adjusted by introducing a right-handed neutrino
superfield at an intermediate scale with a soft mass
$m_{\tilde{\nu}_{R}}^{2} < 0$.  A new neutrino Yukawa term
$h_{\nu}^{2}m_{\tilde{\nu}_{R}}^{2}$ then enters the RGE for $m_2^2$,
and may be used to balance the RGE.  The additional freedom results from
the fact that $m_{\tilde{\nu}_{R}}^{2}$ is unconstrained, as the
physical mass of the scalar neutrino is determined essentially by the
intermediate scale.

\section{The $CP$ Problem, \rb, and Proton Decay}

The light Higgsino-gaugino window has a number of interesting
consequences.  With conventional weak scale SUSY breaking parameters,
the present bound on the electric dipole moments (EDMs) of atoms,
molecules, and the neutron limit the $CP$ violating phases in the
dimensionful parameters of the MSSM to be less than $10^{-2} - 10^{-3}$
over much of the parameter space \cite{edmbounds}.  This is generally
referred to as the SUSY $CP$ problem.  As shown in Ref. \cite{relax},
all flavor-conserving $CP$ odd observables are proportional to the
phases of $M_{\lambda} \mu (m_{12}^2)^*$, $A^* M_{\lambda}$, or $A \mu
(m_{12}^2)^*$, where $M_{\lambda}$ is any one of the three gaugino
masses.  The phase of $M_3$ does not enter the electron EDM at one-loop.
It follows that the electron EDM is proportional at lowest order to at
least one insertion of $\mu$, $M_1$, or $M_2$.  For $\mu, M_1, M_2
\sim {\cal O}({\rm GeV})$, one see that the electron EDM is suppressed by
${\cal O}(M/M_{\mbox{\tiny SUSY}})$, where $M \sim {\cal O}({\rm GeV})$.
This largely eliminates the SUSY $CP$ problem for atoms with unpaired
electrons which are sensitive to the electron EDM.  In addition, if
leptonic $A$ terms are small, as would be the case if the approximate
$U(1)_R$ symmetry discussed above were extended to the leptonic sector,
the electron EDM would be suppressed by ${\cal O}(M/M_{\mbox{\tiny
SUSY}})^2$.

$CP$ violation in the strongly interacting sector depends on the gluino
mass $M_3$, and so is not necessarily suppressed in the
phenomenologically allowed window.  However, if the approximate $U(1)_R$
is extended to the entire Lagrangian, then all gaugino masses, $\mu$,
and all $A$ terms are suppressed.  The EDMs of the neutron and atoms
with paired electrons (which are senstive to strong sector $CP$
violation) are then suppressed by ${\cal O}(M/M_{\mbox{\tiny
SUSY}})^2$.\footnote{The full $U(1)_R$ symmetry imposed in
Refs.~\cite{HallRandall,RR,RS,bw} is not required to solve the SUSY $CP$
problem.  If any two types of the four classes of dimensionful
parameters $\{M_{\lambda}, \mu, A, m_{12} \}$ are ${\cal O}(M) \ll
M_{\mbox{\tiny SUSY}}$ at the high scale, then all $CP$ odd observables
are suppressed by ${\cal O}(M/M_{\mbox{\tiny SUSY}})^2$.  This may be
verified by noting that $U(1)_{PQ}$ and $U(1)_{R-PQ}$ field
redefinitions \cite{relax} may be used to isolate the phases on the
small parameters.}  Even in the RSB scheme given in the previous section
with a large gluino mass and small $A$ terms at the high scale, the
strong sector $CP$ violation is still suppressed by ${\cal
O}(M/M_{\mbox{\tiny SUSY}})$.  This is apparent for the first and third
type of combinations of $CP$ violating parameters given above, as both
involve an insertion of $\mu$.  For the second type this follows since,
even though a sizeable $A$ term can be induced by the gluino from
running to the low scale, the phase is then aligned with that of the
gluino mass, {\it i.e.}, ${\rm Arg}(A) \simeq {\rm Arg} (M_3)$.

Supersymmetric models can in principle give large enough one-loop
corrections to the $Z^0 b \bar{b}$ vertex to explain the $\sim 3\sigma$
discrepancy between the experimental \cite{Rbexp} and Standard Model
values of $R_b$ \cite{finnell,Rbrefs,Wagner,Jorge}.  The most important
contributions are from vertex corrections involving a top quark Yukawa
coupling $b_L \tilde{H}_2 \tilde{t}_R$.  A sizeable effect requires a
light chargino with a substantial Higgsino component, $\tan
\beta \simeq 1$, and a light $\tilde{t}_R$
\cite{finnell,Rbrefs,Wagner,Jorge}. The first two of these requirements
are met in the light Higgsino-gaugino window.  In addition, as
demonstrated earlier, it is also possible to arrange for a light
$\tilde{t}_R$ consistent with RSB.  This is highly non-trivial, as it is
generally difficult to obtain solutions that explain the $R_b$
discrepancy consistent with RSB.  (See, however, Ref.~\cite{Wagner} for
a solution with conventional weak-scale SUSY parameters.)  The effect in
the light Higgsino-gaugino window (see, for example, Ref.~\cite{Jorge}),
may be determined from the figures of Ref.~\cite{finnell}.  We find that
for $\tan \beta = 1$, $\mu = M_2 = 0$, and
$m_{\tilde{t}_R} = 100$ GeV, the SUSY shift in $R_b$ is $\delta R_b
\approx 0.002$, or roughly equal to what can be achieved in the Higgsino
region with similar chargino and stop masses.\footnote{Here we have
ignored left-right stop mixing angle suppressions.  In principle, the
mixing, which in our case is due to weak scale $A$ parameters, can be
small for $m_{\tilde{t}_R} \lesssim m_t$ if the soft mass squared
$m_{\tilde{t}_R}^2$ is slightly negative at the weak scale (when
permitted by the stability of the potential).}  This effect is, of
course, greatly increased for smaller $m_{\tilde{t}_R}$, and may
therefore significantly reduce (but not eliminate) the current
discrepancy of 0.006 between experiment and the Standard Model
\cite{Rbexp}.

Proton decay at one-loop is also suppressed in the allowed window.  The
supersymmetric baryon violating coupling $QQQL$ must be dressed with an
off-shell gaugino in order to obtain a four-Fermi interaction.  With
degenerate squarks, and ignoring any flavor changing, the gluino
contribution vanishes, so the largest dressing typically comes from
charginos \cite{proton}.  A chiral insertion is necessary on the
chargino line to obtain the four-Fermi interaction.  In order to avoid a
light quark Yukawa coupling to the Higgsino component of the chargino,
an $M_2$ insertion is required.  The proton decay rate is then
suppressed in this limit at one-loop by ${\cal O} (M_2/M_{\mbox{\tiny
SUSY}})^2 \sim 10^{-4}$ in the allowed window.  If the gluino is
massive, gluino exchange could then dominate the decay rate if there are
flavor changing squark masses.

\section{Hidden Sector Scenarios}

Finally, let us consider a possible theoretical motivation for the light
Higgsino-gaugino window.  In hidden sector models, SUSY breaking is
transmitted to the visible sector by gravitational strength
interactions.  With a renormalizable hidden sector, in which SUSY
breaking remains in the flat space limit, the dynamical scale,
$\Lambda$, of the hidden sector gauge group, and the hidden sector
scalar expectation values, $Z$, are of the order of the intrinsic SUSY
breaking scale, $M_S \sim \Lambda \sim Z \sim \sqrt{m_{3/2} M_p} \sim
10^{10-11}$ GeV.  This allows an expansion of the operators which couple
the visible and hidden sectors in powers of $M_p^{-1}$.  In the rigid
supersymmetric limit, the dimension two soft terms arise from $D$ term
operators of the form ${1 \over M_p^2} \int d^4 \theta ~Z^*Z \phi^*\phi$
and ${1 \over M_p^2} \int d^4 \theta ~Z^*Z H_1 H_2$ \cite{gm}, where $Z$
are any hidden sector fields and $\phi$ is a visible sector field.  With
$F_Z \sim M_S^2$, we then have $m_{\phi}^2 \sim m_{12}^2 \sim
m_{3/2}^2$.  Note that these dimension two terms arise even without
hidden sector singlets.  The dimension three gaugino masses arise from
the dependence of a visible sector gauge kinetic function on a hidden
sector singlet $S$, which does not transform under any gauge symmetry,
${1 \over M_p} \int d^2 \theta ~S W^{\alpha} W_{\alpha}~+\mbox{ h.c.}$
Visible sector $A$ terms arise from $D$ term operators ${1 \over M_p}
\int d^4 \theta ~S \phi_i^*\phi_i~+\mbox{ h.c.}$, where $F_{\phi_i}^* =
h_{ijk} \phi_j \phi_k$ results from the visible sector Yukawa couplings
$W = h_{ijk} \phi_i \phi_j \phi_k$.\footnote{It is interesting to note
that the resulting $A$ terms are real. Independent of their magnitude,
$A$ terms therefore do not contribute to the SUSY $CP$ problem with a
renormalizable hidden sector.}  Likewise, the $\mu$ term arises from
operators of the form ${1 \over M_p} \int d^4 \theta ~S H_1 H_2~+ \mbox{
h.c.}$ \cite{gm}.

If the hidden sector singlets have $F$ components $F_S \sim M_S^2$, then
all the dimensionful parameters of the MSSM can be ${\cal O}(m_{3/2})$.
However, it is possible that the hidden sector singlets participate in
the supersymmetry breaking only radiatively so that $F_S \sim {\cal
O}(\lambda/4 \pi)^2 M_S^2$, where $\lambda$ is a hidden sector singlet
Yukawa coupling.  All the dimension three soft terms and $\mu$ are then
automatically suppressed by ${\cal O}(\lambda/ 4 \pi)^2$.\footnote{The
$\mu$ term can also arise from an $H_1 H_2$ dependence of a hidden
sector gauge kinetic function ${1 \over M_p^2} \int d^2 \theta~ H_1 H_2
(W^{\alpha} W_{\alpha}) \vert_{\mbox{\tiny hidden}}~+\mbox{ h.c.}$
\cite{gaugemu}.  For a renormalizable hidden sector with $\langle
W^{\alpha} W_{\alpha} \rangle \sim \Lambda^3 \sim M_S^3$ the resulting
$\mu$ term is very small.  However, a non-renormalizable hidden sector
with $\langle W^{\alpha} W_{\alpha} \rangle \sim \Lambda^3 \sim M_S^2
M_p$ gives $\mu \sim m_{3/2}$.  The scenario discussed by Farrar and
Masiero in which all the dimension three terms but $\mu$ essentially
vanish \cite{lightgluino} is therefore realizable with a
non-renormalizable hidden sector without singlets and with scalar
expectation values much less than $M_p$.  The magnitude of the $\mu$
term therefore distinguishes between the renormalizable and
non-renormalizable hidden sector motivations for light gauginos.}
Inclusion of supergravity interactions does not modify this conclusion.
The smallness of the dimension three terms in such a scenario leads to
the approximate $U(1)_R$ discussed above.  This approximate symmetry is
not imposed by hand but simply arises accidentally as a result of the
hidden sector outlined above.  Notice that this motivation for the
window requires the gluino also to be light
\cite{lightgluino,moregluino}. Models with radiatively coupled
singlets have in fact been constructed \cite{bkn} and until recently
were the only known renormalizable models of dynamical supersymmetry
breaking with singlets \cite{quantum}.  We therefore conclude that a
renormalizable hidden sector with radiatively coupled singlets
automatically leads to models that can fall in the light
Higgsino-gaugino window.

\section{Conclusions}

At present there exists a small region of supersymmetric parameter space
in which $\mu$ and the electroweak gaugino masses $M_1$ and $M_2$ are in
the few GeV range.  This window has a number of interesting
consequences: 1) The SUSY $CP$ problem can be significantly reduced, 2)
the discrepancy in $R_b$ can be substantially reduced if the
right-handed stop is light, and 3) proton decay is suppressed.  The mass
of the lightest Higgs is generated almost entirely radiatively since
$\tan \beta \simeq 1$, and requires a fairly heavy stop to exceed
current bounds. (This must be the left-handed stop if a light
right-handed stop is required for $R_b$.)  Radiative electroweak
symmetry breaking is generally difficult in the allowed window with
minimal particle content, but can be accommodated.  In particular, a
heavy gluino or intermediate scale right-handed neutrino can allow
radiative symmetry breaking.  Renormalizable hidden sectors with
radiatively coupled singlets automatically give models that can fall in
the allowed window.  Most importantly, this window predicts that two
neutralino states are light, two are roughly degenerate with the $Z^0$,
and both charginos are roughly degenerate with the $W^{\pm}$.  All of
these particles cannot escape detection at LEP II.

\acknowledgements

We would like to thank G. Farrar, H. Haber, L. Hall, J. Louis, H.
Murayama, H.--P.  Nilles, M. Peskin, S. Pokorski, and M. Srednicki for
useful discussions.  We also thank D. Finnell, J. Lopez, C. Wagner, and
J. Wells for discussions of numerical results for $R_b \,$.
N. P. acknowledges the theoretical physics group of the Lawrence
Berkeley National Laboratory for hospitality while some of this work was
completed.

\figure{\label{fig:1}
Allowed regions of the $(\mu, M_2)$ plane for $\tan\beta = 1.15$ and
$M_1/M_2 = \frac{1}{2}$ (solid), 1 (dashed), and 2 (dotted).  These
regions satisfy the bounds of Ref.~\cite{L3} from data taken at
$\sqrt{s} = M_Z$.  }

\figure{\label{fig:2}
The region (shaded) of the $(\mu, M_2)$ plane that satisfies the refined
bounds from radiative photon decays and data taken 1.8 GeV above $M_Z$
(see text).  Here $\tan\beta = 1.15$, and only points that satisfy the
bounds for some $M_1$ in the range $\frac{1}{2} M_2 \le M_1 \le 2 M_2$
are considered allowed.  For $M_1>M_2$, the region can be extended to
lower $M_2$ for $\mu<0$.}

\figure{\label{fig:3}
Typical spectra found for the light and pseudoscalar Higgs bosons, top
and bottom squarks with dominant right- and left-handed components, and
the gluino, using the boundary conditions described in the text and
requiring RSB.  The Higgs boson is constrained to be heavier than 60
GeV.
%%\postscript{mu0fig3.ps}{1}
}

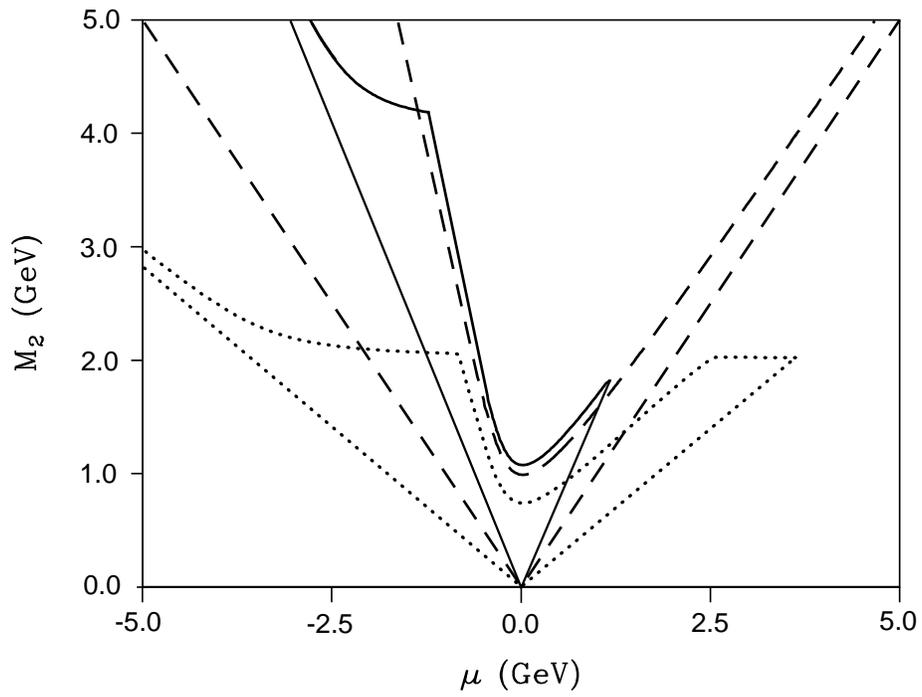
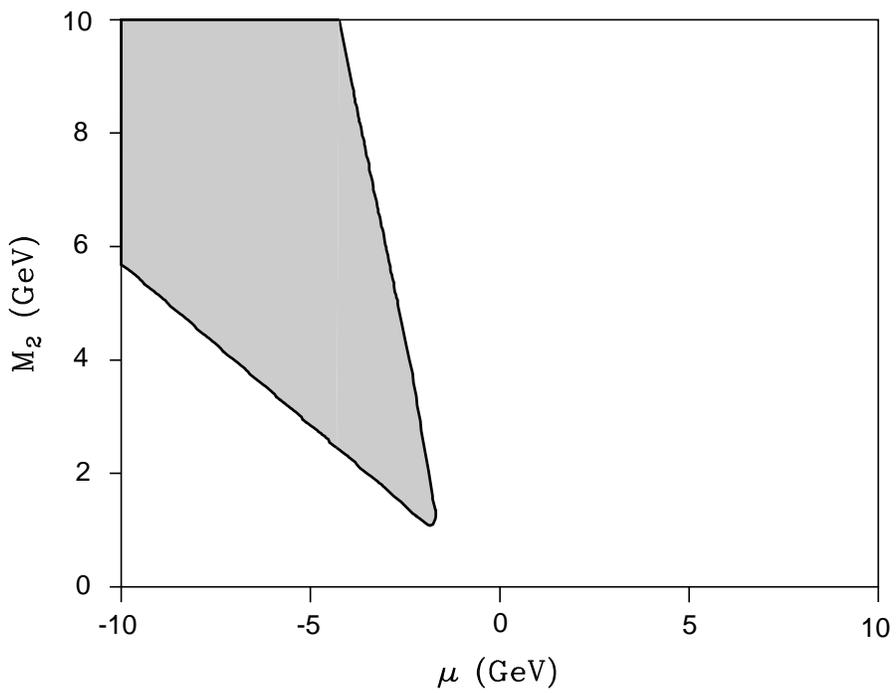
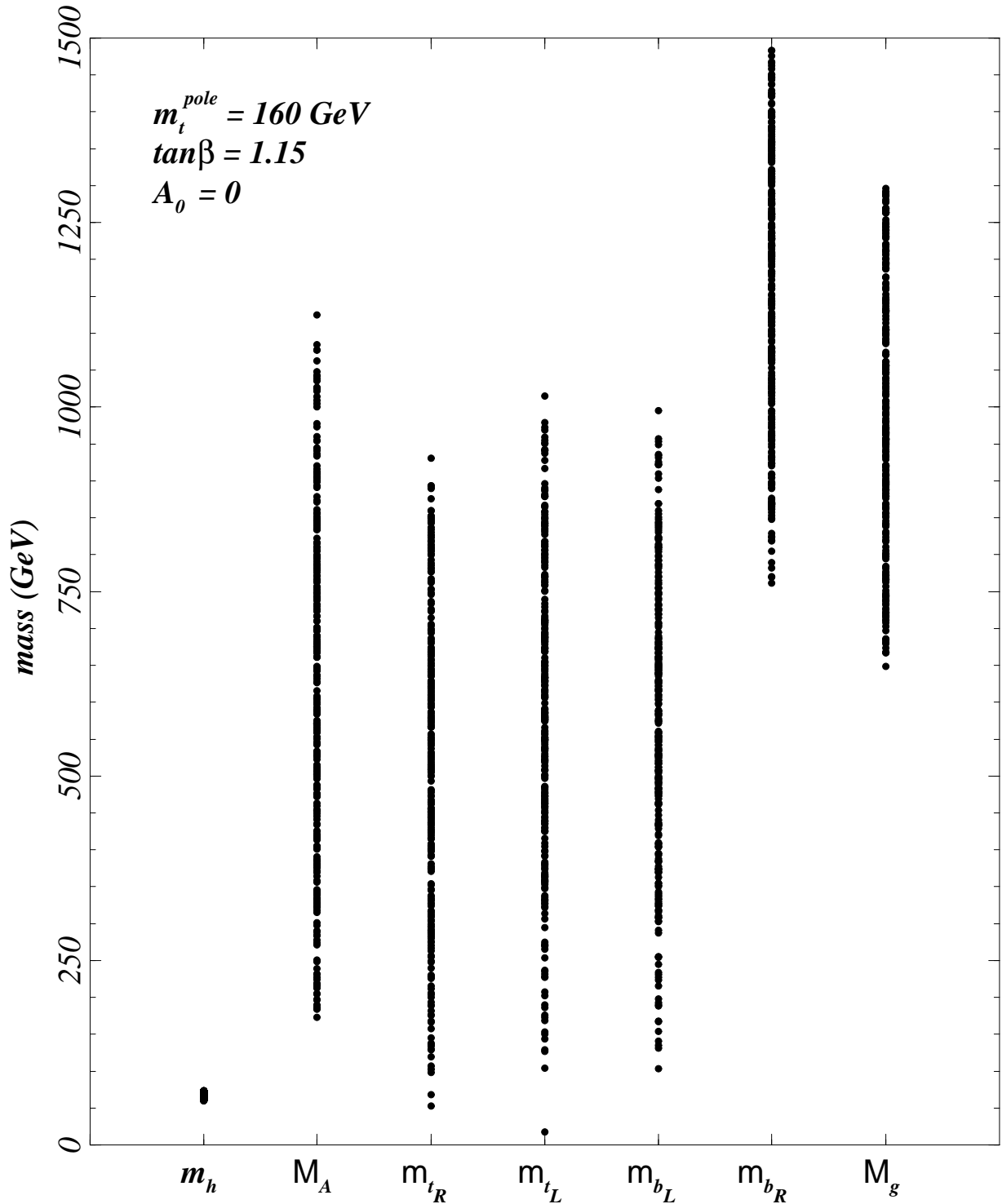
\end{document}